# Thermochemical Stability of Low-Iron, Manganese-Enriched Olivine in Astrophysical Environments

Denton S. Ebel[1,2,4*], Michael K. Weisberg[1,3,4], John R. Beckett[5]

[1]Department of Earth and Planetary Sciences, American Museum of Natural History, New York, NY 10024.
[2]Department of Earth and Environmental Sciences, Lamont-Doherty Earth Observatory of Columbia University, New York, NY.
[3]Department of Physical Sciences, Kingsborough College of CUNY, 2001 Oriental Blvd., Brooklyn, NY 11235.
[4]Graduate Center of the City University of New York, NY.
[5]Division of Geological and Planetary Sciences, California Institute of Technology, Pasadena, CA 91125.
*Corresponding author: debel@amnh.org

**Abstract** - Low-iron, manganese-enriched (LIME) olivine grains are found in cometary samples returned by the Stardust mission to comet 81P/Wild 2. Similar grains are found in primitive meteoritic clasts and unequilibrated meteorite matrix. LIME olivine is thermodynamically stable in a vapor of solar composition at high temperature at total pressures of a millibar to a microbar, but enrichment of solar composition vapor in a dust of chondritic composition causes the FeO/MnO ratio of olivine to increase. The compositions of LIME olivines in primitive materials indicate oxygen fugacities close to that of a very reducing vapor of solar composition. The compositional zoning of LIME olivines in amoeboid olivine aggregates is consistent with equilibration with nebular vapor in the stability field of olivine, without reequilibration at lower temperatures. A similar history is likely for LIME olivines found in comet samples and in interplanetary dust particles. LIME olivine is not likely to persist in nebular conditions in which silicate liquids are stable.

# INTRODUCTION

A condensed solar composition with all Fe as FeO has a FeO/MnO wt% ratio of 95.5 (Anders and Grevesse 1989). Low-iron, manganese-enriched (LIME) olivine with far lower FeO/MnO ratios are, however, a ubiquitous component of some of the most primitive materials formed in the solar system (Fig. 1). These include cometary particles from the Stardust sample suite (Zolensky et al. 2006; Nakamura et al. 2008), chondritic interplanetary dust particles (IDPs, Klöck et al. 1989; Nakamura-Messenger et al. 2010), meteoritic amoeboid olivine aggregates (AOAs, Grossman and Steele 1976; Aléon et al. 2002; Weisberg et al. 2004; Sugiura et al. 2009), chondrule olivines in CR chondrites (Ichikawa and Ikeda 1995), and olivine in the matrices of primitive chondrites such as Semarkona (LL3.0) and Murchison (CM2; Klöck et al. 1989). LIME low-Ca pyroxene is also present in Stardust samples (Zolensky et al. 2006) and in the Allende (CV3) chondrite (Rubin 1984). LIME silicates (primarily olivine) have wt% FeO/MnO less than and, usually, much less than, 10 and low FeO, typically with <1.0 wt% FeO, which corresponds to $X_{Fa}$ <0.01, where $X_{Fa}$ is the mole fraction fayalite ($Fe_2SiO_4$).

The compositions of LIME olivines and pyroxenes are significantly different from the compositions typical of ferromagnesian silicates from undifferentiated meteoritic material. For the latter, MnO contents of olivine and pyroxene are almost always below 0.5 wt%, or $X_{Tep}$ < 0.005, where $X_{Tep}$ is the mole fraction tephroite ($Mn_2SiO_4$) in olivine, with no significant correlation between FeO and MnO contents (Klöck et al. 1989, their Fig. 1b). MnO contents of olivine in FeO-poor chondrules of CR chondrites do increase slightly with increasing FeO content, but most of the wt% FeO/MnO ratios in olivine remain above 5 (Weisberg et al. 1995).

Here, we explore the effect of oxygen fugacity on the relative stabilities of olivine components in this composition range. We present the results of thermochemical calculations, which suggest that LIME silicates are the natural outcomes of high-temperature equilibration between solids and vapors expected in protoplanetary disks and other astrophysical environments for a restricted suite of bulk chemical compositions. Thus, the existence of LIME silicates places important constraints on environments of formation, and silicate crystallization in these environments. We note that, although these calculations may apply to other astrophysical environments (e.g., atmospheres of AGB stars), the focus here is on nebular environments of olivine formation.

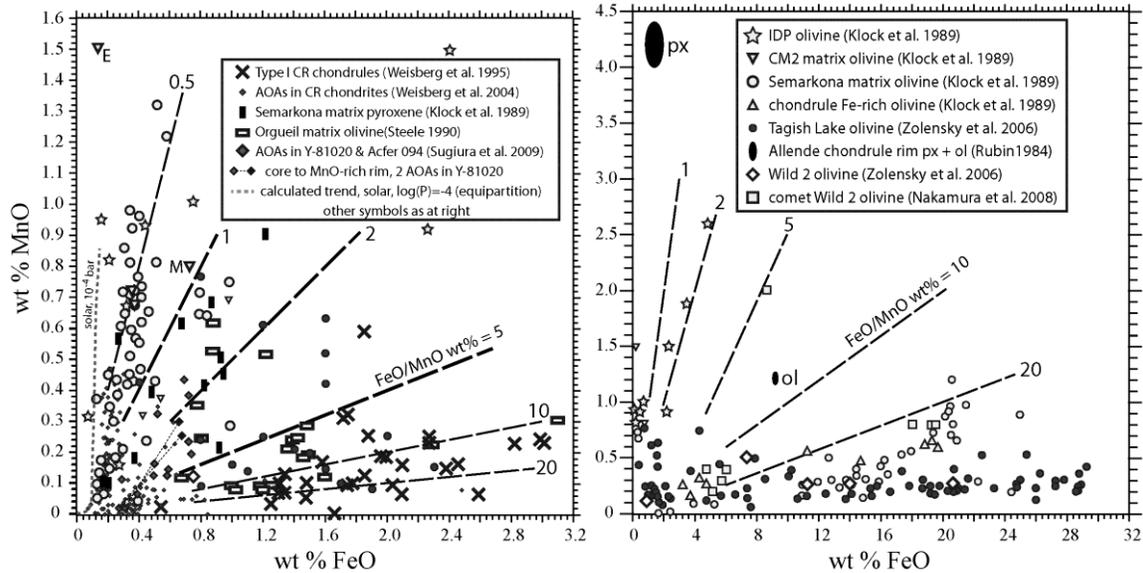

**Fig. 1:** Compositions of olivine in primitive solar system materials. Panel "**a**" shows detail in the dashed rectangle of panel "**b**". Reported MnO contents of meteoritic and IDP olivines are plotted as a function of wt% FeO, where mole fractions tephroite ($X_{Tep}$) and fayalite ($X_{Fa}$) are nearly equal to wt% MnO/100 and wt% FeO/100, respectively, at the low $X_{Tep}$ and $X_{Fa}$ of "**a**". Long dashed lines indicate constant wt% FeO/MnO ratios. Data are from Klöck et al. (1989, their Table 1 and Fig. 1), Weisberg et al. (1995, their Fig. 6; 2004), Zolensky et al. (2006, their Fig. S2b), Rubin (1984, his Table 2, px: pyroxene, ol: olivine), Steele (1990, his Tables 1,3), Sugiura et al. (2009, their Table 1), and Nakamura et al. (2008, their Table S1). Matrix olivines in "**a**" are from CM2 chondrites (E: EET83226, and M: Murchison; inverted triangles) and Orgueil (CI; open rectangles). Two populations of Semarkona (LL3.0) matrix olivines (open circles) are FeO-rich, in "**a**", and MnO-rich, in "**b**" (Klöck et al. 1989). Core and MnO-rich rim analyses of olivine from two AOAs (Sugiura et al. 2009) are indicated. The dotted line in "**a**" indicates the condensation trend for a vapor of solar composition at $P^{tot} = 10^{-4}$ bar, with equipartition of Mn between pyroxenes and olivine (see Fig. 3).

## METHODS I: CONDENSATION CALCULATIONS

A straightforward approach to understanding LIME olivine stability is to incorporate a complete olivine solid solution model directly into the VAPORS code of Ebel and Grossman (2000; cf. Ebel et al. 2000; Ebel 2006), and to then explore olivine stability at fixed total pressure, $P^{tot}$, and total composition as a function of temperature. In a condensation calculation, the free energy of the system is minimized to determine the stable phase assemblage and the composition of each phase at equilibrium. Ebel and Grossman (2000) only allowed for (Mg, Fe, Ca)$_2$SiO$_4$ olivine solid solutions. A more recent thermodynamic model that is now embedded in MELTS accounts for the solution of additional divalent transition elements in olivine, including Mn (M. Hirschmann, pers. comm.; Sack and Ghiorso 1989; Hirschmann 1991; Hirschmann and Ghiorso 1994). We have incorporated this model into the VAPORS code so that the stability and compositions of olivines with the general formula $(Mg,Fe,Ca,Mn,Co,Ni)_2SiO_4$ can be evaluated. Since VAPORS is based on the MELTS thermodynamic crystallization package (Ghiorso and Sack, 1995), internal consistency with the implementation of this olivine model is assured. Sugiura et al. (2009) presented calculations of wt% MnO in olivine in equilibrium with a vapor of solar composition at $P^{tot} = 10^{-4}$ bar. They allowed Mn condensation only into olivine (cf., Petaev and Wood 2005). For comparison, we also calculate compositions of olivine solid solution in equilibrium with a cooling vapor of solar composition at $P^{tot} = 10^{-4}$ bar.

In the VAPORS code, $H_2$-rich vapor, silicate liquid, olivine, MnS, the tephroite ($MnTiO_3$) component of rhombohedral oxide (Ghiorso 1990), $Mn^o$ (metal), MnSi (brownleeite), and MnO are the only phases allowed to contain Mn. A variety of Mn-bearing gaseous species are considered but, for all conditions investigated here, Mn in the vapor phase speciates overwhelmingly as monatomic $Mn_{(g)}$. Mn solubility in pyroxenes is not included in our calculations, even though LIME pyroxenes exist in primitive meteorites. Robust equations of state for Mn-bearing multicomponent pyroxenes, consistent with the MELTS database, do not yet exist, as noted by Petaev and Wood (2005). However, we do provide a zeroth order assessment based on the known similarity in partitioning for Mn among olivine and pyroxenes (Kennedy et al. 1993; http://earthref.org/KDD/e:25/). Furthermore, the calibration of $(Mg, Fe, Ca, Mn, Co, Ni)_2SiO_4$ olivine solid solution properties must be extrapolated beyond its range of calibration for some of the calculations reported here. Nominal maximum temperatures of calibration are 1370 K for fayalite, 1500 K for tephroite, and 1807 K for forsterite. Finally, we note that grains of low-iron, chromium-enriched olivine (LICE-ol) are also observed in some samples from comet 81P/Wild 2 (Zolensky et al. 2006), and in some primitive meteorites and IDPs (Klöck et al. 1989). Sugiura et al. (2009) did calculate the stability of the Cr component in forsteritic olivine, using a model by Li et al. (1997). Our analysis does not extend to Cr-bearing olivine compositions, for lack of an activity model for Cr-olivine that is consistent with the model of Hirschmann and Ghiorso (1994).

## METHODS II: ELLINGHAM DIAGRAMS

An alternative springboard for exploration of the Mn, Cr and Fe content of silicates is the Ellingham diagram (Ellingham 1944). These diagrams, familiar in the metallurgical literature, describe the relative oxidation/reduction or sulfidation/reduction tendencies of pure metals and compounds at smelting temperatures T by plotting the Gibbs free energies of oxidation (or sulfidation) reactions, where $\Delta G^o_{rxn} = RT\ln(p_{O_2})$. Richardson and Jeffes (1948) developed this concept into a graphical representation of metal-oxide equilibrium relations in T - $RT\ln(p_{O_2})$ space from which $CO/CO_2$, $H_2/H_2O$, and $p_{O_2}$ could be read. This 'nomogram' allowing analog interpolation for a myriad of buffer curves from a single diagram is a classic example of compact, interpretable presentation of information (Tufte 2001). Examples of this nomogram are given by Darken and Gurry (1953, their Fig. 14-4), and by Larimer (1967, his Fig. 4) in illustrating the effect of oxidation potential on oxides in ordinary and enstatite chondrites. We will use this diagram to explore the relative stabilities of olivine components.

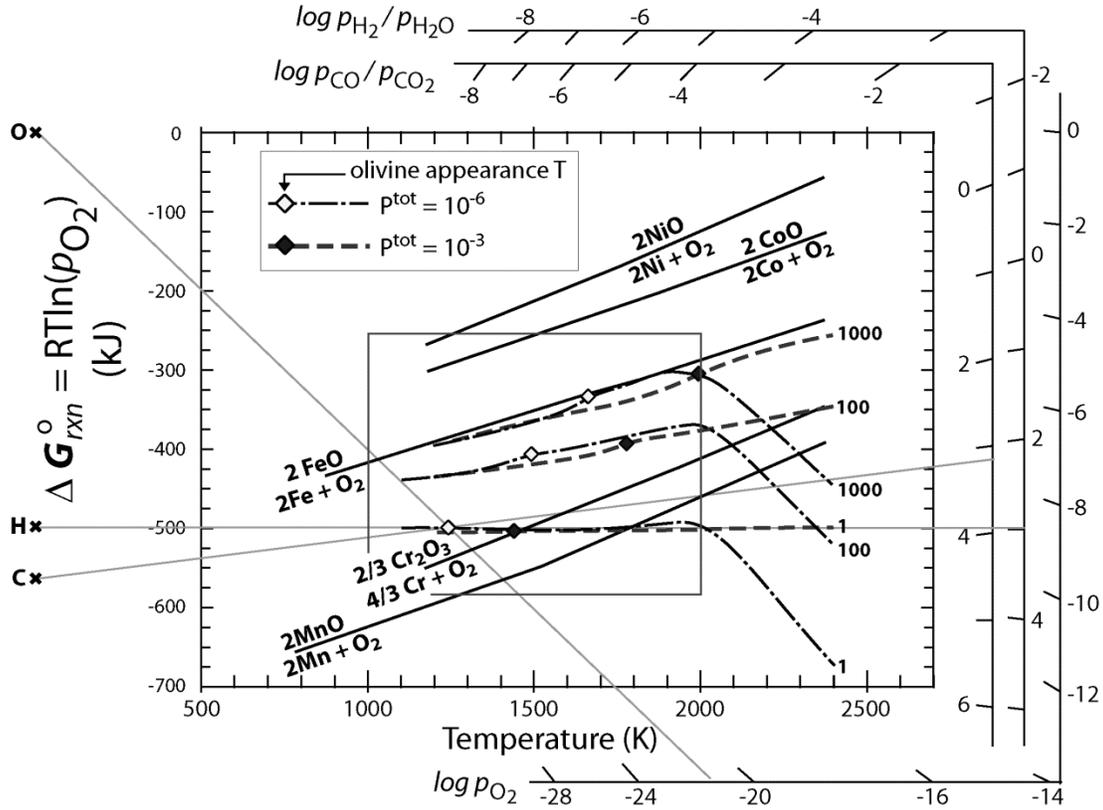

**Fig. 2:** Ellingham diagram showing metal-oxide buffer curves. For these equilibria the vertical axis represents the Gibbs free energy of reactions, $\Delta G_{rxn}$, written with $O_{2(gas)}$ as a reactant, which is quantitatively equivalent to $RT \ln(p_{O_2})$. Changes in slope correspond to changes in state of the metals and oxides. Superimposed on the metal-oxide equilibria are $p_{O_2}$ trajectories of cooling vapors of solar composition (labeled 1), and for solar compositions that are enriched with dust of CI bulk composition by a factor of 100 (labeled 100) and 1000 (labeled 1000) calculated for $P^{tot} = 10^{-3}$ (dashed curves) and $P^{tot} = 10^{-6}$ bar (dot-dash curves). The initial condensation temperature of olivine for each of these curves is indicated by a filled ($P^{tot} = 10^{-3}$) or open ($P^{tot} = 10^{-6}$) diamond. All of the curves shown in the panel were obtained using the current VAPORS code (Ebel and Grossman 2000; Ebel 2006; this work). Inset rectangle is the focus of Figs 4 and 5. Outside the central panel is a set of Richardson-Jeffes style nomogram scales that can be used to determine the *log* of $p_{O_2}$, $p_{H_2}/p_{H_2O}$ and $p_{CO}/p_{CO_2}$ at any desired point within the panel. For example, a line of constant $H_2/H_2O$ ratio can be constructed by drawing a line segment originating at the **x** labeled H to the left of the diagram (this point plots at 0 K) through a point of interest within the panel (e.g., olivine-in for condensation of a system of solar composition at $P^{tot} = 10^{-6}$ bar) and extending it to the scale labeled *log* $(p_{H_2}/p_{H_2O})$, which reads 3. Every point along this line segment has the same $p_{H_2}/p_{H_2O}$ ratio, $10^3$. Similar lines can be constructed for $p_{O_2}$, drawing from the **x** labeled O, and $p_{CO}/p_{CO_2}$, drawn from the **x** labeled C.

Figure 2 follows the diagram of Darken and Gurry (1953). Plotted are reaction equilibria for oxides of selected metals (Mn, Cr, Fe, Co, and Ni) calculated with VAPORS (Ebel and Grossman 2000). The oxide is stable above each curve. Also shown are $T - RT\ln(p_{O_2})$ trajectories for equilibrium condensation of solids from a vapor of solar composition (Anders and Grevesse 1989) at $P^{tot} = 10^{-3}$ and $10^{-6}$ bar, calculated for the temperature range 2400 - 1150 K (Ebel 2006). These curves are labeled "1" to indicate the bulk composition of the system has 1 times CI abundances of the condensable elements. At lower temperatures, both of these curves converge to $p_{H_2}/p_{H_2O} \approx 10^3$. At high temperatures and low $P^{tot}$, the $p_{O_2}$ of these chemical systems is not dominated by the reaction $2 H_2 + O_2 = 2 H_2O$, so that the $P^{tot} = 10^{-6}$ trajectory dips below $p_{H_2}/p_{H_2O} \approx 10^3$.

In addition to the solar condensation trajectories shown in Fig. 2, condensation paths for vapors that have been enriched in dust of CI chondrite composition relative to solar by factors of 100 and 1000 (Ebel and Grossman 2000) are also shown. It is reasonable to expect enrichment of the solar nebula in previously condensed dust in the disk midplane, in the region of terrestrial planet formation. A major effect of enrichment in a carbonaceous chondrite-like (CI) dust is to decrease the $H_2/H_2O$ ratio of the system, because dust brings substantial oxygen in the form of condensed silicates and little hydrogen. Note for a given enrichment in dust, trajectories for $P^{tot} = 10^{-6}$ and $10^{-3}$ bar diverge at high temperatures, as they do for the solar case, and converge at low temperatures to a specific $p_{H_2}/p_{H_2O}$ dictated by the degree of dust enrichment.

Relative positions of oxidation equilibria in Fig. 2 give a general idea of the propensity for a given element for oxidized or reduced forms. Thus, we would expect Ni and Co, under the relatively reducing conditions considered here, to be generally in the metallic phase, Cr and Mn to be oxidized and Fe to vary depending on specific conditions. In detail, however, it is the solution of an element into major silicates like olivine and pyroxene that dictates the distribution. For Mn, it is the stability of the Mn-endmember (tephroite) in olivine that is of interest. At any particular temperature and $P^{tot}$, the reaction

$$2Mn_{(g)} + O_{2(g)} + SiO_{(g)} = Mn_2SiO_{4(in\ olivine)} \qquad (1)$$

may be described by a Gibbs free energy of reaction ($\Delta G_{rxn}$):

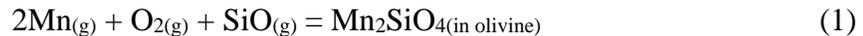
$$\Delta G_{rxn} = G^o_{Mn_2SiO_4} + RT\ln a^{olivine}_{Mn_2SiO_4} - 2G^o_{Mn(g)} - 2RT\ln p_{Mn} - G^o_{O_2(g)} - RT\ln p_{O_2} - G^o_{SiO(g)} - RT\ln p_{SiO} \qquad (2)$$

where $R$ is the gas constant (Jmol$^{-1}$K$^{-1}$), $T$ is temperature (Kelvin), $a^{olivine}_x$ is the activity of endmember $x$ (e.g., tephroite) in olivine, $G^o_s$ are the standard state Gibbs free energies of species $s$, and $p_y$ is the partial pressure of gas species $y$ in the vapor. Equation 2 is valid at low pressures for which fugacities are equivalent to partial pressures of gaseous species.

For equation 2, and its analog for fayalite, $\Delta G_{rxn} = 0$ at equilibrium. Upon rearranging the equation, $RT\ln(p_{O_2})$ can be obtained for fixed mole fractions of olivine endmembers, over a range of temperatures:

$$RT \ln p_{O_2} = G^o_{Mn_2SiO_4} + RT \ln a^{olivine}_{Mn_2SiO_4} - 2G^o_{Mn(g)} - 2RT \ln p_{Mn} - G^o_{O_2(g)} - G^o_{SiO(g)} - RT \ln p_{SiO} \quad (3)$$

This requires calculation of $a^{olivine}_{Mn_2SiO_4}$, partial pressures of gaseous species, and the $G^o_s$, at each particular temperature. Here, we use this calculation as a basis for exploring the relative stabilities of olivine with fixed mole fractions of tephroite or fayalite in solid solution, by reference to the Ellingham diagram (Fig. 2). The activities $a^{olivine}_{Mn_2SiO_4}$ and $a^{olivine}_{Fe_2SiO_4}$ in olivine are calculated for fixed mole fractions of 0.001, 0.01, and 0.025 at each temperature using the models described by Sack and Ghiorso (1989) and Hirschmann and Ghiorso (1994), a set of compositions that spans most of the observed range of LIME olivines (Fig. 1a). Values for the partial pressures of $SiO_{(g)}$, $Mn_{(g)}$ and $Fe_{(g)}$ in equation 2, and in the analogous equations for fayalite, are taken directly from condensation calculations at the appropriate total composition, temperature, and $P^{tot}$ (Ebel and Grossman 2000). An implicit assumption in this approach is that small degrees of solid solution of tephroite or fayalite in olivine do not affect the activity of the other endmember in the same solid solution. Values for $G^o_s$ are calculated using equations of state for olivine endmembers and gaseous species (Hirschmann and Ghiorso 1994; Ebel and Grossman 2000). In essence, the calculations establish the redox conditions under which olivine of a particular composition can be in equilibrium with a particular gas.

We focus below on condensation trajectories of a vapor of solar composition (Anders and Grevesse 1989) at $P^{tot}$ in the range $10^{-3}$ to $10^{-6}$ bar, and several dust enriched compositions, over the temperature range 2400 - 1150 K taken at 10 degree steps. These calculated trajectories yield partial pressures of SiO ($p_{SiO}$) and other gaseous species at each step. Because equation 2, and also the condensation calculations, describes chemical equilibrium conditions, the gas species used in equation 2 are inconsequential. For example, although $SiO_{(g)}$ is the dominant Si-bearing species in solar and more oxidizing vapors, equations equivalent to equation 2 could be written that incorporate alternative species and this would have no effect on the calculated Mn concentrations in olivine.

## RESULTS

Figure 3 illustrates the result of a direct condensation calculation (Method I) for the solids, including multicomponent olivine, from a vapor of solar composition at $P^{tot} = 10^{-4}$ bar (10 Pa). Results for $P^{tot}$ from $10^{-2}$ to $10^{-6}$ bar are qualitatively similar, because, although changes in $P^{tot}$ shift the temperatures of solid stability fields, the relative shifts are similar for each crystalline phase. In these calculations, mole fractions of Mn endmember (tephroite, Tep) in olivine peak at 0.025, when ~100% of the Mn has condensed. Since a vapor of canonical solar composition has considerably more Mg and Fe than Mn (atomic ratios of Mn/Mg = 0.009, and Mn/Fe = 0.011; Anders and Grevesse 1989), the condensation of MnO into olivine has a negligible effect on condensate modes or appearance temperatures.

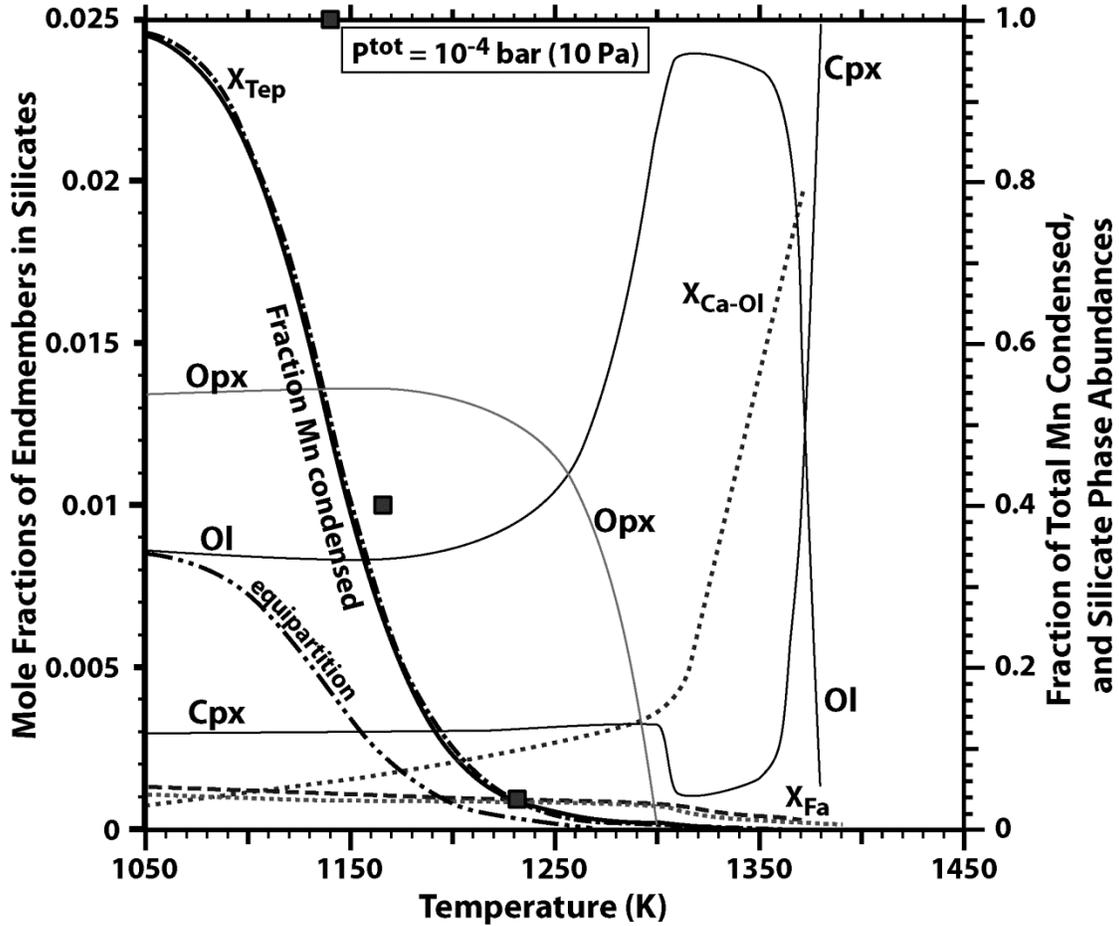

**Fig. 3:** End-member mole fractions, and relative abundances, of silicates calculated to condense from a vapor of solar composition at $P^{tot} = 10^{-4}$ bar as a function of temperature. Olivine is the only silicate that VAPORS allows to dissolve Mn, and the fraction of total Mn in the system condensed as olivine in this calculation is indicated by the solid black curve. The corresponding mole fraction of tephroite in the olivine ($X_{Tep}$, dash-dot line) reaches a high of nearly 0.025. The curve "XCa-Ol" (dotted) is the calculated mole fraction of a $Ca_2SiO_4$ component in olivine. Values of $X_{Fe-Opx}$ (not shown) are virtually equal to $X_{Fa}$ (dashed line) below 1300 K. Fractional molar abundances of olivine (Ol), Ca-free orthopyroxene (Opx), and Ca-rich clinopyroxene (Cpx) are relative to the sum of Ol+Opx+Cpx. Using molar distribution coefficients for Mn of unity among these three phases and assuming that the calculated amount of Mn condensed into olivine represents the total amount of Mn condensed, results in the Mn endmember mole fractions indicated by the dot-dot-dash curve labeled "equipartition". Temperatures calculated for $X_{Tep} = 0.001$, 0.01, and 0.025 using method II (see text) are indicated by filled squares.

    A weakness of the current VAPORS code is that partitioning of Mn among coexisting olivine, Ca-rich pyroxene (clinopyroxene, Cpx), and Ca-free pyroxene (orthopyroxene, Opx) is not considered. We address this by assuming an $Mn_2Si_2O_6$ endmember for both pyroxenes, and Mn partition coefficients among olivine and the two pyroxenes of unity (equipartition). We assume that the affinity of the Mn endmember in

the pyroxenes is the same as in olivine, and, therefore, assume that the calculated olivine Mn-component abundance represents the total condensable Mn at each temperature. Results of this calculation are shown in Fig. 3. The mole fraction of the Mn endmember in all three silicates reaches a maximum of 0.0085 (i.e., 1/3 of the maximum for olivine computed using the VAPORS code) or ~0.85 wt% MnO. Most of the LIME olivines plotted in Fig. 1a fall below this value.

Direct condensation calculations account for the depletion of Mn in the vapor at temperatures below the appearance of olivine. Equation 3 can be used to calculate $p_{O_2}$ for fixed $X_{Tep}$ under the assumptions that $p_{Mn}$ and $p_{SiO}$ are constant. The first assumption becomes less tenable as the vapor becomes more depleted in Mn, a factor not accounted for in the source calculations for the vapor pressures used in equation 2. Thus, the temperature at which $X_{Tep} = 0.001$ according to Method II and equation 2 is essentially equivalent to that obtained from a full-scale condensation calculation, but the temperatures at which $X_{Tep} = 0.01$ and 0.025 are reached are higher by 15 and 90 K, respectively, at $P^{tot} = 10^{-4}$ bar.

In addition to showing $X_{Tep}$ as a function of temperature, Fig. 3 also illustrates the variation of $X_{Fa}$ and the influence of condensate metal on Fe in coexisting silicates. Forsteritic olivine condenses at 1370 K from a vapor of solar composition at $P^{tot} = 10^{-4}$ bar, and metal alloy condenses at 1360 K, significantly lowering $p_{Fe}$ in the vapor. Mn is unaffected because of a negligible solubility in the alloy. Condensate olivines quenched at high temperature have $X_{Fa} > X_{Tep}$, but both are extremely low. As temperature decreases, Mn in olivine increases more rapidly than Fe and, below 1240 K, $X_{Tep} > X_{Fe}$. However, little new olivine or pyroxene condenses under these conditions (Fig. 3). Overall, these calculations imply that a condensate olivine exposed to a vapor of solar composition over a range of temperatures during cooling, that is not reequilibrated will be normally zoned (i.e., Mn will increase from core to rim).

A strength of Method II is that it allows a simple assessment of the influence of changing redox conditions on the compositions of phases. Figure 4 illustrates the result of applying equation 3 (Method II) to a portion of Fig. 2, for a single condensation scenario. Calculated isopleths are plotted for fixed mole fractions of tephroite and fayalite components dissolved in olivine solid solutions condensing from a vapor of solar composition at $10^{-3}$ bar, calculated using values for $p_{SiO}$, $p_{Mn}$, and $p_{Fe}$ from condensation calculations (Ebel and Grossman 2000), at those chemical and pressure conditions, at each temperature. For modest perturbations from the solar condensation curve, values of $p_{SiO}$, $p_{Mn}$, and $p_{Fe}$ change much more slowly than $p_{O_2}$, and we can, therefore, use Fig. 4 to deduce relative changes in the olivine composition caused by changes in the redox conditions. Note that the intersection of a tephroite isopleth with the solar condensation trajectory occurs at a higher temperature than the intersection for a fayalite isopleth of the same mole fraction. Thus, at 1300 K, olivine containing $X_{Tep} = 0.001$ and $X_{Fa}$ ~0.001 is stable, but $X_{Tep}$ increases to 0.025 at 1200 K whereas $X_{Fa}$ remains ~0.001.

In more oxidizing systems, for example systems enriched in ices, the trajectories must shift to higher $p_{O_2}$ in the Ellingham diagram. Thus, the addition of water ice to a system of solar composition will stabilize fayalite relative to tephroite, at the temperatures considered here. The addition of chondritic dust to a vapor of solar composition would cause a similar shift in oxidation potential of the system, and preferentially stabilize the fayalitic component of olivine. For example, Fig. 4 illustrates the $p_{O_2}$ trajectory of a vapor 1000x enriched in CI chondrite dust. Below 1600 K, the corresponding trajectory for a vapor 50x enriched in CI chondritic dust would plot subparallel to the $X_{Fa} = 0.01$ curve. Perturbations in gas composition that lead to conditions more reducing than those of a solar gas (e.g., removal of ices, or C enrichment) would have the opposite effect and destabilize fayalite relative to tephroite.

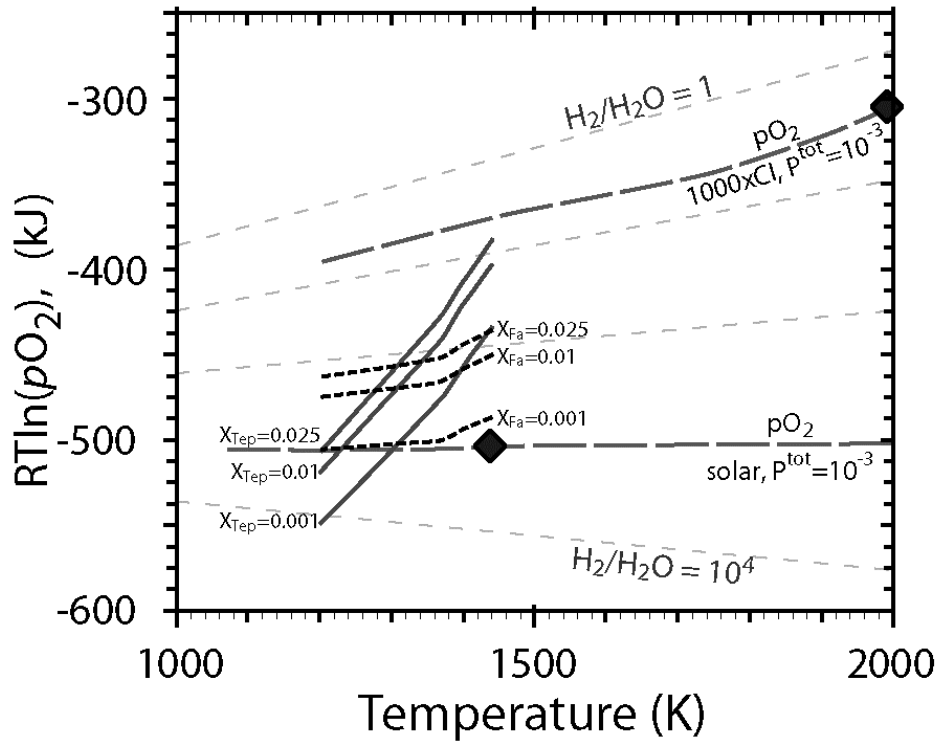

**Fig. 4:** Isopleths of the tephroite (Tep, solid curves) and fayalite (Fa, short dash curves) components in olivine, in a vapor of solar composition at $P^{tot} = 10^{-3}$ bar. Mole fractions 0.001 (lowest $\Delta G_{rxn}$ at a given temperature), 0.01, and 0.025 (highest $\Delta G_{rxn}$) of components are plotted. Also shown are $p_{O_2}$ trajectories of cooling vapor (dashed), with initial condensation temperature of olivine marked (diamond), for the solar, $P^{tot} = 10^{-3}$ bar case (1444 K), and also for a vapor enriched 1000x in CI-type dust at $P^{tot} = 10^{-3}$ bar (1990 K). Light dashed lines show constant $\log p_{H_2}/p_{H_2O}$ ratios of 0, 1, 2, and 4 (see Fig. 2).

Figure 5 illustrates the effects of varying $P^{tot}$ and dust enrichment on olivine composition. Stability curves for 0.001, 0.01, and 0.025 mole fraction tephroite and fayalite in olivine are shown. Values for the partial pressures of $SiO_{(g)}$, $Mn_{(g)}$ and $Fe_{(g)}$ in equation 3 used to calculate these curves were taken directly from condensation calculations at the appropriate total composition, T, and $P^{tot}$. For all $10^{-6} \leq P^{tot} \leq 10^{-2}$ bar, Mn-enriched olivine is more thermodynamically stable than Fe-enriched olivine in a cooling gas of solar composition at temperatures within the olivine stability field. Figure 5 also illustrates the effect of 100x enrichment of the system in a dust of CI composition, relative to solar, at constant $P^{tot} = 10^{-3}$ bar. Ebel and Grossman (2000; cf., Ebel 2006) demonstrated that silicate liquids similar to those quenched in chondrules become stable at dust enrichments between ~15xCI and 1000xCI at $P^{tot} = 10^{-3}$ bar. At 100xCI dust enrichment, olivine incorporates the fayalite component of a given mole fraction at higher temperature than it does the tephroite component. When $X_{Tep}$ reaches 0.001, olivine already contains $X_{Fa} > 0.01$. At 100x enrichment, a nearly constant ~0.025 mole fraction fayalite is stabilized, and primary olivine ceases to condense below ~1620 K (cf., Ebel and Grossman 2000, their Fig. 8a). Were olivine to continue to equilibrate with vapor to lower temperatures, mole fractions of tephroite would increase, reaching $X_{Tep}$ ~ $X_{Fa}$ ~ 0.025 at 1500 K. At 1000xCI dust enrichment, olivine condensation ceases at ~1680 K, and the fayalite component is much more stable than the tephroite component in olivine at all temperatures. Therefore, $X_{Tep} > X_{Fa}$ only at dust enrichments well below 100xCI. From Fig. 5, it is clear that decreasing $H_2/H_2O$ will decrease the Mn/Fe ratio in olivine for any particular $P^{tot}$ or dust enrichment condition.

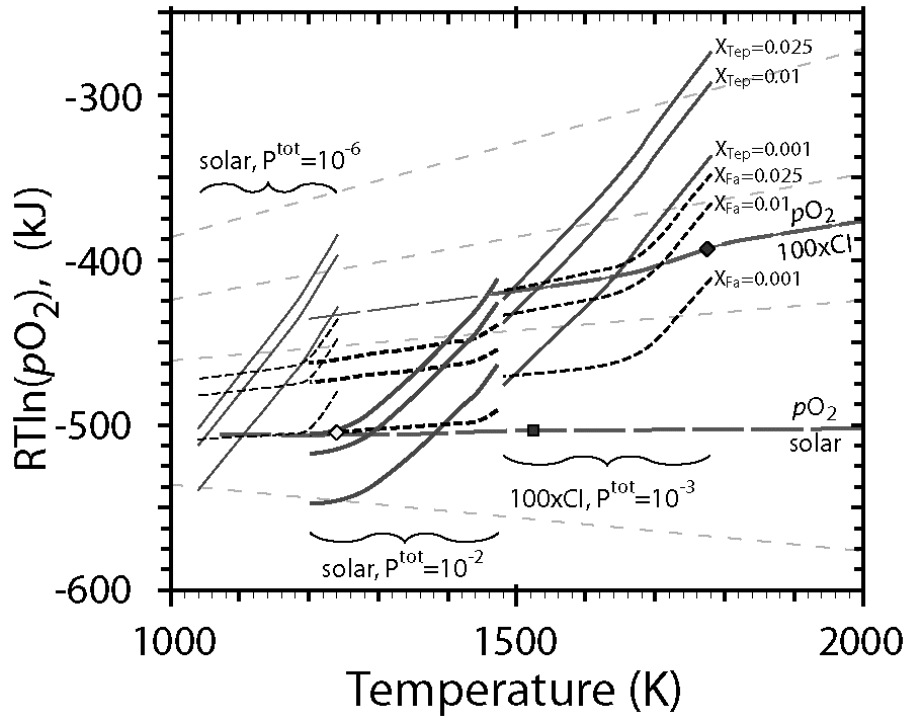

**Fig. 5:** Effect of total pressure ($P^{tot}$) and dust enrichment on tephroite (Tep, solid curves) and fayalite (Fa, short dashes) components in olivine. The condensation trajectory for a vapor of solar composition (curves for $P^{tot} = 10^{-6}$ to $10^{-2}$ bar are indistinguishable on the scale of this figure) and of a vapor enriched 100x in CI-type dust at $P^{tot} = 10^{-3}$ bar are shown as long dashed curves. In a vapor of solar composition, olivine appears at 1240 K at $P^{tot} = 10^{-6}$ bar (open diamond) and 1520 K at $P^{tot} = 10^{-2}$ bar (filled square); olivine appears at 1780 K for 100xCI $P^{tot} = 10^{-3}$ bar (filled diamond). Select lines of constant $log\ p_{H_2}/p_{H_2O}$, taken from Fig. 4, are also shown.

**DISCUSSION**

From Fig. 3 and the analysis of Figs. 4 and 5 above, Mn concentrations in forsteritic olivine condensed from solar compositions, and from solar compositions enriched by CI dust, are expected to increase with decreasing temperature. This is qualitatively consistent with olivine in AOAs, which are Mn-poor in AOA interiors and Mn-enriched in their rims (Weisberg et al. 2004; Sugiura et al. 2009). Although Ca, Ti, and Al can also substitute into the olivine crystal structure at high temperatures (Steele 1990; Spandler and O'Neill 2010), they condense more readily into oxide and silicate solids at these conditions, and their affinity for olivine is low at the temperatures where Mn concentrations in olivine reach their maxima. For the conditions explored here, Mn (and Cr) are only incorporated into pyroxenes coexisting with olivine. Nakamura-Messenger et al. (2010) reported brownleeite (MnSi) epitaxially intergrown with LIME olivine in an anhydrous chondritic-porous IDP, probably from comet 26P/Grigg-Skjellerup. Here, brownleeite was found to be unstable in a solar composition vapor at $P^{tot} = 10^{-4}$ bar at any temperature, suggesting that even more reducing conditions are required to stabilize this phase. Although we are not able to effectively model Cr condensation into olivine or pyroxene, consideration of Fig. 2 suggests that Cr should enter silicates less readily than Mn under the same redox conditions (e.g., $H_2/H_2O$ ratio). Sugiura et al. (2009), however, found that the CrO (wt%) content of olivine exceeded MnO above ~1200 K. They suppressed formation of Cr-rich spinel, which is not found in association with AOAs, and would become stable at 1200 K in their calculation. In the present calculation, Cr- spinel (~$Mg_{1.03}Al_{0.77}Cr_{1.16}Ti_{0.04}O_4$) becomes stable at 1220 K. The elements Ni and Co, which readily enter the olivine structure under favorable conditions, require much more oxidizing conditions than Mn to do so; for the astrophysically plausible redox conditions explored here, these elements will be strongly reduced and concentrated in coexisting metal alloy.

    The condensation behavior of Mn has previously been reported in detail only by Sugiura et al. (2009), with markedly different results from the present calculation. Larimer (1967) calculated the condensation temperature of pure tephroite ($Mn_2SiO_4$) as 1470 K in a vapor with solar oxygen fugacity at $P^{tot} = 6.6 \times 10^{-3}$ bar at ~1100 K. Larimer (1973, his Fig. 2b) calculated a 50% condensation temperature ($T_c$) for Mn as $MnSiO_3$ of ~1250 K at $P^{tot}=10^{-4}$ bar; Wai and Wasson (1977, their Table 1) calculated $T_c$ as 1190 K for the same conditions, with activity coefficients of $Mn_2SiO_4$ of unity. Palme and Fegley (1990) predicted FeO/MnO ratios in olivine for high $H_2O/H_2$ ratios, neglecting the details of gas depletion in Fe and Mn and their incorporation into other phases. Ebel and Grossman (2000), and Petaev et al. (2003) did not include Mn in silicate solid solutions. Lodders (2003) calculated the $T_c$ for Mn as 1158 K in a vapor of solar composition at $P^{tot}=10^{-4}$ bar, with Mn entering both olivine and orthopyroxene solid solutions, using methods described by Kornacki and Fegley (1986).

    The preservation of LIME olivines in a wide range of primitive nebular materials (Fig. 1) indicates that these objects did not undergo re-equilibration with Fe-bearing vapor or condensed material under oxidizing conditions. Such re-equilibration would have dramatically enriched the olivine in Fe. The coexistence of LIME olivine with LIME pyroxene (e.g., in cometary grains, Nakamura et al. 2008) suggests that these phases formed co-genetically. However, the data of Nakamura et al. (2008) allow calculation of apparent distribution coefficients (wt% MnO in olivine/wt% MnO in pyroxene) ranging from 0.95 to 4.0, so a variety of local environments may have been sampled. In a cooling vapor of solar composition, enstatite forms by the reaction of forsteritic olivine with Si from vapor (Grossman 1972; Ebel and Grossman 2000).

The formation of LIME olivine is highly sensitive to redox conditions. In the chemical systems investigated here, LIME olivine is only stable in reduced vapors, with CI dust enrichments well below 100x. There is, therefore, a very limited range of CI dust enrichments, below ~50x, in which both silicate melts and LIME olivines may be stable under the same $P^{tot}$ and temperature conditions. For similar reasons, the objects in which LIME silicates are found cannot have formed in regions enriched in water ice. LIME olivine is primarily found in unmelted objects such as IDPs, AOAs, and primitive chondrite matrices. These objects show little evidence for pervasive melting under conditions where silicate melt is stable against evaporation (Ebel 2006). Mn-enrichment should be observed at the outer edges of primitive (i.e., highly magnesian), unequilibrated olivine and pyroxene grains formed under low $p_{O_2}$ conditions. The chondrule-like objects recovered from comet Wild 2 (Nakamura et al. 2008) are not particularly Mn-enriched (Fig. 1b), consistent with the fact that conditions that stabilize silicate liquids do not stabilize LIME olivine.

*Acknowledgements* -- The authors thank two reviewers. This work was supported by NASA-COS-NNX10AI42G (DSE), NASA-COS-NNG09AG94G (MKW), NASA-DDAP-NNX08AG23G (MKW), and NASA-COS-NNX09AG40G (E. Stolper, PI). This research has made use of NASA's Astrophysics Data System